\documentclass[12pt]{article}

\usepackage{graphicx}
\usepackage{amsmath}

\def\ltwid{\mathrel{\raise.3ex\hbox{$<$\kern-.75em\lower1ex\hbox{$\sim$}}}}
\def\gtwid{\mathrel{\raise.3ex\hbox{$>$\kern-.75em\lower1ex\hbox{$\sim$}}}}

\def\square{\kern1pt\vbox{\hrule height 1.2pt\hbox{\vrule width 1.2pt\hskip 3pt
   \vbox{\vskip 6pt}\hskip 3pt\vrule width 0.6pt}\hrule height 0.6pt}\kern1pt}
\def\overleftrightarrow#1{\vbox{\ialign{##\crcr
     $\leftrightarrow$\crcr\noalign{\kern-1pt\nointerlineskip}
     $\hfil\displaystyle{#1}\hfil$\crcr}}}

\begin{document}

\begin{titlepage}

\begin{flushright}
UFIFT-QG-12-07
\end{flushright}

\vskip 1cm

\begin{center}
{\large\bf Representing the Vacuum Polarization on de Sitter}
\end{center}

\vskip .5cm

\begin{center}
Katie E. Leonard$^{1*}$, Tomislav Prokopec$^{2\dagger}$ 
and Richard P. Woodard$^{1,\ddagger}$
\end{center}

\vskip .5cm

\begin{center}
\it{$^{1}$ Department of Physics, University of Florida \\
Gainesville, FL 32611, UNITED STATES}
\end{center}

\begin{center}
\it{$^{2}$ Institute of Theoretical Physics (ITP) \& Spinoza Institute, \\
Utrecht University, Postbus 80195, 3508 TD Utrecht, \\ THE NETHERLANDS}
\end{center}

\vskip .5cm

\begin{center}
ABSTRACT
\end{center}

\noindent
Previous studies of the vacuum polarization on de Sitter have 
demonstrated that there is a simple, noncovariant representation
of it in which the physics is transparent. There is also a 
cumbersome, covariant representation in which the physics is
obscure. Despite being unwieldy, the latter form has a powerful 
appeal for those who are concerned about de Sitter invariance. 
We show that nothing is lost by employing the simple, noncovariant 
representation because there is a closed form procedure for 
converting its structure functions to those of the covariant
representation. We also present a vastly improved technique for
reading off the noncovariant structure functions from the 
primitive diagrams. And we discuss the issue of representing the
vacuum polarization for a general metric background.

\begin{flushleft}
\end{flushleft}

\vskip .5cm

\begin{flushleft}
$^*$ e-mail: katie@phys.ufl.edu \\
$^{\dagger}$ e-mail: T.Prokopec@uu.nl \\
$^{\ddagger}$ e-mail: woodard@phys.ufl.edu
\end{flushleft}

\end{titlepage}


\section{Introduction}
\label{Introduction}

The vacuum polarization $i[\mbox{}^{\mu} \Pi^{\nu}](x;x')$ is a 
bi-vector density which can be used to quantum-correct the 
classical Maxwell equations,
\begin{equation}
\partial_{\nu} \Bigl[ \sqrt{-g} \, g^{\nu\rho} g^{\mu\sigma} 
F_{\rho\sigma}(x) \Bigr] + \int \!\! d^4x' \Bigl[ \mbox{}^{\mu}
\Pi^{\nu} \Bigr](x;x') A_{\nu}(x') = J^{\mu}(x) \; . \label{maxeqn}
\end{equation}
The vacuum polarization encodes two sorts of information:
\begin{itemize} 
\item{How quantum 0-point fluctuations affect photons; and}
\item{How quantum 0-point fluctuations modify the electromagnetic
response to charges and currents.}
\end{itemize}
In flat space background, neither charged matter nor gravitons have
any effect on photons, and significant modifications to the 
electromagnetic response are limited to the immediate vicinity of 
the source \cite{Peskin:1995ev,Leonard:2012fs}. 
An analogous local effect occurs for gravitons~\cite{Marunovic:2012pr}.
Things can be very
different in de Sitter background. The inflationary production of
light, minimally coupled, charged scalars induces so much vacuum
polarization that the photon develops a mass
\cite{Prokopec:2002jn,Prokopec:2002uw,Prokopec:2003iu,Prokopec:2003tm}
and electromagnetic forces suffer comparable changes \cite{Degueldre:2012}.
A study of the effect of inflationary gravitons is far advanced 
\cite{Leonard:2012}. Although they cannot induce a photon mass, 
it is expected that buffeting by the vast ensemble of infrared gravitons
will lead to a progressive growth of the field strength. This occurs
for massless fermions \cite{Miao:2005am,Miao:2006gj,Miao:2007az}, and 
probably also for massive ones \cite{Miao:2012bj}, owing to the 
spin-spin coupling \cite{Miao:2008sp}, which photons also possess.

In view of its physical importance, consideration should clearly be given 
to the best way of representing the tensor structure of $i[\mbox{}^{\mu} 
\Pi^{\nu}](x;x')$. It is transverse on each index, and also symmetric
under interchange,
\begin{equation}
i \Bigl[ \mbox{}^{\mu} \Pi^{\nu}\Bigr](x;x') =
i \Bigl[ \mbox{}^{\nu} \Pi^{\mu}\Bigr](x';x) \; .
\end{equation}
These two facts mean $i[\mbox{}^{\mu} \Pi^{\nu}](x;x')$ can have at 
most $\frac12 \times 4 \times 5 - 4 = 6$ independent components. 
However, on flat space background the vacuum polarization can be
represented using only a single structure function,
\begin{equation}
i \Bigl[ \mbox{}^{\mu} \Pi^{\nu}_{\rm flat}\Bigr](x;x') =
-\Bigl[ \eta^{\mu\nu} \partial^2 \!-\! \partial^{\mu} \partial^{\nu}
\Bigr] \Pi( \Delta x^2) \; ,
\end{equation}
where $\eta^{\mu\nu}$ is the spacelike Minkowski metric and
$\Delta x^2 \equiv \Vert \vec{x} - \vec{x}' \Vert^2 - (|x^0 -
{x'}^0 | - i \epsilon)^2$. 

On the homogeneous and isotropic backgrounds of cosmology the
vacuum polarization can be written as a linear combination of
four algebraically independent tensors \cite{Miao:2010vs}. 
Transversality provides two differential relations between
their coefficients, so there should be two independent
structure functions \cite{Miao:2010vs} in a general cosmological
background. The earliest de Sitter computations
\cite{Prokopec:2002jn,Prokopec:2002uw,Prokopec:2003tm} were made 
in conformal coordinates,
\begin{equation}
ds^2 = a^2 \Bigl( -d\eta^2 + d\vec{x} \!\cdot\! d\vec{x} \Bigr) 
\qquad , \qquad a(\eta) = -\frac1{H \eta} \; ,
\end{equation}
where $H$ is the Hubble constant. These early works represented 
the vacuum polarization as,
\begin{equation}
i \Bigl[ \mbox{}^{\mu} \Pi^{\nu}\Bigr](x;x') =
\Bigl[ \eta^{\mu\nu} \eta^{\rho\sigma} \!-\! \eta^{\mu\sigma} 
\eta^{\nu\rho} \Bigr] \partial_{\rho} \partial'_{\sigma} F(x;x')
+ \Bigl[ \overline{\eta}^{\mu\nu} \overline{\eta}^{\rho\sigma} 
\!-\! \overline{\eta}^{\mu\sigma} \overline{\eta}^{\nu\rho} \Bigr] 
\partial_{\rho} \partial'_{\sigma} G(x;x') \; , \label{oldway}
\end{equation}
where $\overline{\eta}^{\mu\nu} \equiv \eta^{\mu\nu} + \delta^{\mu}_0
\delta^{\nu}_0$ is the purely spatial part of the Minkowski metric.

The structure functions $F(x;x')$ and $G(x;x')$ have a simple
interpretation in terms of the electric and magnetic susceptibilities 
$\chi_e$ and $\chi_m$ \cite{Prokopec:2003bx},
\begin{equation}
\chi_e(x;x') = -i F(x;x') \qquad , \qquad \frac{\chi_m}{1+\chi_m}(x;x') 
= -i G(x;x') \; .
\end{equation}
These susceptibilities can be used to obtain the polarization and the
magnetization vectors, and the corresponding macroscopic fields,
the same as for flat space electrodynamics in a medium,
\begin{eqnarray}
\vec{P} & = & \chi_e \otimes \vec{E} \qquad \Longrightarrow \qquad
\vec{D} = \vec{E} + \vec{P} \; , \label{polarization} \\
\vec{M} & = & \chi_m \otimes \vec B \qquad \Longrightarrow \qquad
\vec{H} = \vec{B} - \vec{M} \; . \label{magnetization}
\end{eqnarray}
Here $\otimes$ signifies convolution, meaning that 
relations (\ref{polarization}-\ref{magnetization}) are actually non-local 
and can involve integrations within the past light-cone.

Although the original representation~(\ref{oldway}) has a
transparent physical interpretation, and is easy to use in the 
quantum-corrected Maxwell equations~\cite{Prokopec:2003iu}, 
its structure functions $F(x;x')$ and $G(x;x')$ are not bi-scalar 
densities because $\eta^{\mu\nu}$ and $\overline{\eta}^{\mu\nu}$ are 
not bi-tensors. 
 This may appear disturbing to those who believe that   
the de Sitter group should play the 
same role in organizing quantum field theory on de Sitter background
that the Poincar\'e group does on flat space 
\cite{Higuchi:2011vw,Miao:2011ng}. It was difficult to form an 
opinion as to the merit of this view as long as only the noncovariant
representation (\ref{oldway}) had been studied. Therefore, we recently 
recast the old results of SQED (scalar quantum electrodynamics) 
\cite{Prokopec:2002jn,Prokopec:2002uw,Prokopec:2003tm} in a covariant 
form \cite{Leonard:2012si},
\begin{eqnarray}
\lefteqn{ i \Bigl[ \mbox{}^{\mu} \Pi^{\nu}\Bigr](x;x') = \sqrt{-g} \, 
\sqrt{-g'} \, D_{\rho} D'_{\sigma} \Biggl\{ D^{\mu} D^{\prime [\nu} y \, 
D^{\prime \sigma]} D^{\rho} y \!\times\! f_1(y,u,v) } \nonumber \\
& & \hspace{5.5cm} + D^{[\mu} y \, D^{\rho] } D^{\prime [\nu} y \,
D^{\prime \sigma]} y \!\times\! f_2(y,u,v) \Biggr\} \; . \qquad
\label{newway}
\end{eqnarray}
Here $D^{\rho}$ and $D^{\prime \sigma}$ stand for covariant derivative
operators, square brackets denote anti-symmetrization, and the 
variables $y$, $u$ and $v$ are,
\begin{equation}
y(x;x') = a a' H^2 \Delta x^2 \quad , \quad u(x;x') = \ln(a a') 
\quad , \quad v(x;x') = \ln\Bigl( \frac{a}{a'} \Bigr) \; . \label{y}
\end{equation}

The de Sitter invariant length function $y(x;x')$ in (\ref{newway})
is related to the geodetic length $\ell(x;x')$ from $x^{\mu}$ to 
${x'}^{\mu}$ by $y = 4 \sin^2( \frac12 H \ell)$. When the vacuum 
polarization is de Sitter invariant, the structure functions are
independent of $u$ and $v$, and only one structure function is required, 
just as in flat space. This happens for the case of a scalar with 
positive mass-squared \cite{Prokopec:2003tm}. Dependence upon the de 
Sitter breaking terms $u(x;x')$ and $v(x;x')$ arises from de Sitter 
breaking in the massless, minimally coupled scalar propagator 
\cite{Allen:1987tz,Onemli:2002hr,Onemli:2004mb}. 

The new representation (\ref{newway}) is not very illuminating, 
even in the massive case for which it is fully de Sitter invariant
\cite{Leonard:2012si}. Renormalization is unbelievably complicated.
(This is also the case for the closely related graviton self-energy
\cite{Park:2011ww}.) The worst problem is that the correction for 
dynamical photons --- which simply degenerates to a local Proca term 
in the noncovariant representation \cite{Prokopec:2003tm} --- takes 
the form of a surface term at the initial time in the covariant 
representation. (This again happens for quantum corrections to 
dynamical gravitons \cite{Park:2011kg}.) We are used to such surface 
terms being suppressed by powers of the scale factor 
\cite{Onemli:2002hr,Onemli:2004mb}, and sometimes being absorbable 
into corrections of the initial state \cite{Kahya:2009sz},
so it is unsettling to have them {\it dominate} local physics in 
the de Sitter covariant formulation. We successfully resolved the 
apparent paradox \cite{Leonard:2012si}, but the covariant representation 
(\ref{newway}) seems to be a cumbersome and nearly useless way of 
describing the vacuum polarization on de Sitter.

There is no reason to expect the covariant representation (\ref{newway})
to be any more useful or revealing for the coming graviton contribution
to the vacuum polarization \cite{Leonard:2012}. Nor can it be de
Sitter invariant. Just as in flat space background \cite{Leonard:2012fs}, 
the primitive one loop diagrams involve photon and graviton propagators.
The photon propagator is manifestly de Sitter invariant in a de Sitter 
invariant gauge \cite{Tsamis:2006gj}, but the graviton propagator must 
break de Sitter invariance 
\cite{Tsamis:1992xa,Miao:2009hb,Miao:2011fc,Kahya:2011sy,Mora:2012zi}. 
Of course this de Sitter breaking is the reason why there can be
a secular enhancement of the field strength, so it should not be 
viewed as an irritant but rather as the signal of the inflationary
graviton production which motivates the computation.

For all these reasons it seems reasonable to represent the primitive
diagrams using the noncovariant representation (\ref{oldway}).
However, 
mathematically minded literature continue to be powerfully drawn to 
de Sitter invariance~\cite{Higuchi:2011vw,Miao:2011ng}, and one
might expect them to prefer the covariant representation (\ref{newway}).
The purpose of this paper is to demonstrate that nothing is lost
by employing the noncovariant representation~(\ref{oldway}). We
will derive a closed form procedure for converting the noncovariant
structure functions $F(x;x')$ and $G(x;x')$ into the structure
functions $f_1(y,u,v)$ and $f_2(y,u,v)$ of the covariant 
representation (\ref{newway}). Although the derivation is highly 
nontrivial, the final result is quite simple, so mathematical
physicists who prefer the covariant representation can easily
convert to it.

Section~\ref{Known Results on Tensor Structures} reviews some
important results which are employed in the analysis. The derivation 
is made in section~\ref{The Derivation}. In 
section~\ref{Correspondence} we check the final transformation
formulae using the vacuum polarization of SQED for which the structure 
functions of both representations are known. We also give a major 
simplification in the procedure for inferring the noncovariant 
structure functions from primitive diagrams. Our conclusions 
comprise section~\ref{Discussion}. After summarizing results
we discuss the fascinating issue of how to represent the vacuum 
polarization on a general metric background.

\section{Known Results on Tensor Structures}
\label{Known Results on Tensor Structures}

In this section we review some recently derived result that are
necessary for the derivation of section~\ref{The Derivation}. The first 
of these results consists of expanding the Minkowski metric and 
its spatial restriction in terms of covariant derivatives of 
$y(x;x')$ and $u(x;x')$ \cite{Kahya:2011sy,Mora:2012kr},
\begin{eqnarray}
\eta^{\mu\nu} \!\!\!\! & = & \!\!\!\! 
\frac{a a'}{2H^2} \Bigl\{ -D^{\mu} D^{\prime \nu} y \!+\!
     D^{\mu} y \, D^{\prime \nu} u \!+\!
       D^{\mu} u \, D^{\prime \nu} y \!-\!
     y D^{\mu} u \, D^{\prime \nu} u \Bigr\} \; , \label{dSeta} \\
\overline{\eta}^{\mu\nu} \!\!\!\! & = & \!\!\!\! 
\frac{a a'}{2H^2} \Bigl\{ -D^{\mu} D^{\prime \nu} y \!+\!
     D^{\mu} y \, D^{\prime \nu} u \!+\!
       D^{\mu} u \, D^{\prime \nu} y \!+\!
     (2 \!-\! y) D^{\mu} u \, D^{\prime \nu} u \Bigr\} \; . \qquad
\label{dSetabar}
\end{eqnarray}
(Note that covariant derivative of $v(x;x')$ are not independent
because $D^{\mu} v = D^{\mu}u$ and $D^{\prime \nu} v = -D^{\prime \nu} u$.)
The second class of results is more involved and concerns the 
most general form for the vacuum polarization \cite{Leonard:2012si},
\begin{equation}
i \Bigl[\mbox{}^{\mu} \Pi^{\nu}\Bigr](x;x') = \sqrt{-g} \,
\sqrt{-g'} D_{\rho} D'_{\sigma} \Bigl[ \mbox{}^{\mu\rho} T^{\nu\sigma}
\Bigr](x;x') \; , 
\label{vacpol:dS}
\end{equation}
where $[\mbox{}^{\mu\rho} T^{\nu\sigma}](x;x') 
= -[\mbox{}^{\rho\mu} T^{\nu\sigma}](x;x') 
= -[\mbox{}^{\mu\rho} T^{\sigma\nu}](x;x') 
= +[\mbox{}^{\nu\sigma} T^{\mu\rho}](x';x)$.

The symmetries of cosmology --- homogeneity and isotropy --- do
not reduce (\ref{vacpol:dS}) to the form (\ref{newway}). In
addition to just $f_1(y,u,v)$ and $f_2(y,u,v)$, the anti-symmetric 
bi-tensor $[\mbox{}^{\mu\rho} T^{\nu\sigma}](x;x')$ can involve
three additional structure functions,
\begin{eqnarray}
\bigl[\mbox{}_{\mu\rho} T_{\nu\sigma} \bigr]
=\partial_\mu \partial'_{[\nu} y \, \partial'_{\sigma]} \partial_\rho y
\!\times\! f_1 + \partial_{[\mu} y \, \partial_{\rho]} \partial'_{[\nu} y \,
\partial'_{\sigma]} y \!\times\! f_2
 + \partial_{[\mu} y \, \partial_{\rho]}
\partial'_{[\nu} y \, \partial'_{\sigma]} u \!\times\! f_3
 \nonumber \\
 +\, \partial_{[\mu}u \, \partial_{\rho]}
\partial'_{[\nu} y \partial'_{\sigma]} y \!\times\! \tilde{f}_3 +
\partial_{[\mu}u \, \partial_{\rho]} \partial'_{[\nu}y \,
\partial'_{\sigma]} u \!\times\! f_4 + \partial_{[\mu}y \, \partial_{\rho]}
u \, \partial'_{[\nu}y \partial'_{\sigma]}u \!\times\! f_5 
\, , \quad
\label{newway2}
\end{eqnarray}
where $f_i=f_i(y,u,v)$ and $\tilde f_3(y,u,v)=f_3(y,u,-v)$, as dictated by 
symmetries of $\bigl[\mbox{}_{\mu\rho} T_{\nu\sigma} \bigr]$. To see that
only two of these structure functions are actually independent we
first note that acting the derivatives in (\ref{vacpol:dS}) results in 
only four algebraically independent tensors,
\begin{eqnarray}
\lefteqn{D^{\rho} {D'}^{\sigma} \bigl[ \mbox{}_{\mu\rho}
T_{\nu\sigma} \bigr](x;x') = \partial_{\mu} \partial'_{\nu} y
\!\times\! F_1 \!+\! \partial_{\mu} y \, \partial'_{\nu} y
\!\times\! F_2} \nonumber \\
& & \hspace{4cm} + \partial_{\mu} u \, \partial'_{\nu} y \!\times\!
F_3 \!+\! \partial_{\mu} y \, \partial'_{\nu} u \!\times\!
\widetilde{F}_3 \!+\! \partial_{\mu} u \, \partial'_{\nu} u
\!\times\! F_4 \, ,
 \qquad \label{Dfs}
\end{eqnarray}
where $F_{i}=F_{i}(y,u,v)$ ($i=1,2,3,4$)
and $\tilde F_3(y,u,v)=F_3(y,u,-v)$. Furthermore, transversality
implies two partial differential relations between the $F_{i}$, one 
proportional to the derivative of $y(x;x')$ and the other proportional 
to the derivative of $u(x;x')$ \cite{Miao:2010vs}.

All of this suggests that there are only two master structure 
functions which we might call $\Phi=\Phi(y,u,v)$ and $\Psi=\Psi(y,u,v)$. 
In ref.~\cite{Leonard:2012si} that suspicion was confirmed by
demonstrating the transversality of the following substitutions for 
the coefficient functions $F_{i}(y,u,v)$,
\begin{eqnarray}
F_1 \!\!\! & = & \!\!\! \Bigl[ -(4 y \!-\! y^2)
\partial_y \!-\! (D\!-\!1) (2 \!-\! y) \!-\! 2 (2 \!-\! y)
\partial_u \!+\! 4 \cosh(v) \partial_u \!-\! 4 \sinh(v) \partial_v
\Bigr] \Phi \nonumber \\
& & \hspace{6.5cm} + \Bigl[4 \cosh(v) \!-\! (2 \!-\! y) \Bigr]
\Psi \!-\! \Xi \; , \qquad \label{F1} \\
F_2 \!\!\! & = & \!\!\! \Bigl[ (2 \!-\! y)
\partial_y \!-\! D \!+\! 1 \!-\! 2 \partial_u \Big] \Phi \!-\! \Psi
\; , \qquad \label{F2} \\
F_3 \!\!\! & = & \!\!\! -2 e^{v} (\partial_u \!-\!
\partial_v) \Phi \!-\! 2 e^{v} \Psi \!+\! \Xi \; , \label{F3} \\
\widetilde{F}_3 \!\!\! & = & \!\!\! -2 e^{-v} (\partial_u \!+\!
\partial_v) \Phi \!-\! 2 e^{-v} \Psi \!+\! \Xi \;
, \label{F3tilde} \\
F_4 \!\!\! & = & \!\!\! -4 \Psi \!+\! (2 \!-\! y) \Xi \; .
\label{F4}
\end{eqnarray}
Here the auxiliary function $\Xi(y,u,v)$ is,
\begin{equation}
\Xi(y,u,v) \equiv (\partial_u^2 \!-\! \partial_v^2) \int \!\! dy \, 
\Phi(y,u,v) \!+\! (2 \!-\! y) \Psi(y,u,v) \!-\! (D \!-\! 2) \!\!
\int \!\! dy \, \Psi(y,u,v) 
\; . 
\label{xi}
\end{equation}
Relations~(\ref{F1}--\ref{F4}) can be inverted to obtain
for the master structure functions,
\begin{eqnarray}
[(D \!-\! 1)  \!+\! 2\partial_u] \Phi &=& -\frac14[(2 \!-\! y) F_1 \!+\!
(4 y \!-\! y^2) F_2 \!+\! (2 \!-\! y) (F_3 \!+\! \widetilde{F}_3)\!-\! F_4]
\qquad 
\label{master function Phi}
\\
 \sinh(v) \Psi
 &=& [\cosh(v) \partial_v \!-\!  \sinh(v) \partial_u] \Phi
-\frac14[F_3 \!-\! \widetilde{F}_3]
 \, . 
\label{master function Psi}
\end{eqnarray}
In the de Sitter invariant case only one 
de Sitter invariant master structure function suffices, 
that is $\Phi = \Phi(y)$ and $\Psi = 0$.

On the other hand, one can relate the master structure functions to 
$f_i$ in~(\ref{newway2}). Below we quote separately 
the contributions from each of $f_i$ to master 
structure functions (this is denoted 
by the subscript $i = 1, 2, 3, 4, 5$ on $\Phi_i$ and $\Psi_i$):
\begin{eqnarray}
\Phi_1 \!\!\!\! & = & \!\!\!\! \frac{H^4}{2} \Bigl\{ (D \!-\! 1) f_1 \!-\! 
(2 \!-\! y) \partial_y f_1 \Bigr\} \; , 
\label{Phi1}
 \\
\Psi_1 \!\!\!\! & = & \!\!\!\! \frac{H^4}{2} \Bigl\{ ( \partial_u^2 \!-\! 
\partial_v^2) f_1 \Bigr\} \; ,
\label{Psi1}
 \\
\Phi_2 \!\!\!\! & = & \!\!\!\! \frac{H^4}{4} \Bigl\{ D (2 \!-\! y) f_2 \!+\! 
(4 y \!-\! y^2) \partial_y f_2 \Bigr\}
 \; , 
\label{Phi2} 
\\
\Psi_2 \!\!\!\! & = & \!\!\!\! \frac{H^4}{4}\Bigl\{ (2 \!-\! y) ( 
\partial_u^2 \!-\! \partial_v^2) f_2 \Bigr\} 
 \; , 
\label{Psi2} 
\\
\Phi_3 \!\!\!\! & = & \!\!\!\! \frac{H^4}{4}  \Bigl\{ -(D \!-\! 1) 
(f_3 \!+\! \widetilde{f}_3) \!+\! (2 \!-\! y)\partial_y (f_3 \!+\! 
\widetilde{f}_3) \!-\! 2  \partial_y(e^{v}
f_3 \!+\! e^{-v} \widetilde{f}_3) \Bigr\}
\; , \qquad
\label{Phi3} 
\\
\Psi_3 \!\!\!\! & = & \!\!\!\! \frac{H^4}{4} \Bigl\{2 \partial_y 
     \Big[\partial_u ( e^{v} f_3 \!+\! e^{-v} \widetilde{f}_3 )
        \!-\! \partial_v ( e^{v} f_3 \!-\! e^{-v} \widetilde{f}_3 )\Big]
 \!-\! (\partial_u^2 \!-\!  \partial_v^2) (f_3
\!+\! \widetilde{f}_3) \Bigr\} 
\,, \qquad
\label{Psi3}
\\
\Phi_4 \!\!\!\! & = & \!\!\!\! \frac{H^4}{4} \Bigl\{ \partial_y f_4 \Bigr\} 
\; , \label{Phi4} 
\\
\Psi_4 \!\!\!\! & = & \!\!\!\! \frac{H^4}{4} \Bigl\{ -(D \!-\! 1) 
\partial_y f_4 \!+\! (2 \!-\! y) \partial_y^2 f_4 \!-\! 2 \partial_y 
\partial_u f_4 \Bigr\} \; .  \qquad \label{Psi4}
\\
\Phi_5 \!\!\!\! & = & \!\!\!\! \frac{H^4}4 \Bigl\{-D f_5 \!+\! 2 (2 \!-\! y)
\partial_y f_5 \!-\! 4 \cosh(v) \partial_y f_5 \Bigr\} 
\; , \label{Phi5} \\
\Psi_5 \!\!\!\! & = & \!\!\!\! \frac{H^4}4 \Bigl\{\! (D \!-\! 1) f_5 \!-\!
(D \!+\! 1) (2 \!-\! y) \partial_y f_5 \!-\! (4 y \!-\! y^2)
\partial_y^2 f_5 \!+\! 2 \partial_u f_5 
\nonumber \\
& & \hspace{0cm} 
+[-\! 2 (2 \!-\! y)\!+\! 4 \cosh(v)] \partial_u\partial_y f_5
\!-\! 4 \partial_v\partial_y [\sinh(v)f_5]
- (\partial_u^2 \!-\! \partial_v^2) f_5 
\Bigr\} 
\; . \qquad 
\label{Psi5}
\end{eqnarray}
These expressions also imply that the vacuum polarization can be
described in terms of any two of the structure functions $f_i(y,u,v)$.
When the result is de Sitter invariant then it requires only a single
structure function, which can be either $f_1(y)$ or $f_2(y)$.


\section{The Derivation}
\label{The Derivation}

The first step in our analysis is to covariantize expression
(\ref{oldway}) by noting it can be written as,
\begin{equation}
  i \Bigl[\mbox{}^{\mu} \Pi^{\nu} \Bigr](x;x')
      = 2 \partial'_{\alpha} \partial_{\beta}
\bigg\{\frac{\sqrt{-g}\sqrt{-g^\prime}}{(aa^\prime)^2}
 \Big[\eta^{\mu[\nu} \eta^{\alpha]\beta} \widehat{F} +
\overline{\eta}^{\mu[\nu} \overline{\eta}^{\alpha]\beta}
\widehat{G} \Big] \bigg\} \,, \label{oldway:2}
\end{equation}
where we have rescaled $F(x;x')$ and $G(x;x')$ as
\begin{equation}
F = \frac{\sqrt{-g}\sqrt{-g^\prime}}{(aa^\prime)^2} \, \widehat{F}
\;,\qquad 
G = \frac{\sqrt{-g}\sqrt{-g^\prime}}{(aa^\prime)^2} \, \widehat{G}
\;.
\label{F G: rescaling}
\end{equation}
Now, when the measure factors $\sqrt{-g}$ and $\sqrt{-g^\prime}$
are pulled out of the curly brackets in~(\ref{oldway:2}), 
the external derivatives become covariant,
\begin{equation}
  i \Bigl[\mbox{}^{\mu} \Pi^{\nu} \Bigr](x;x')
      = 2\sqrt{-g} \sqrt{-g^\prime} \, D'_{\alpha} D_{\beta}
\Big\{(aa^\prime)^{-2}
 \Big[\eta^{\mu[\nu} \eta^{\alpha]\beta} \widehat{F} +
\overline{\eta}^{\mu[\nu} \overline{\eta}^{\alpha]\beta} \widehat{G}
\Big] \Big\} \;. \label{oldway:3}
\end{equation}
This follows from the observation that, because of the anti-symmetry
in $\mu\leftrightarrow\beta$,
$D_{\beta} [^{\mu\beta}T^{\nu\alpha}]=\partial_\beta[^{\mu\beta}T^{\nu\alpha}]
  +\Gamma^\beta_{\beta\sigma}[^{\mu\sigma}T^{\nu\alpha}]$, 
where $\Gamma^\beta_{\beta\sigma}=(-g)^ {-1/2}\partial_\sigma(-g)^ {1/2}$.
The analogous identity holds for the $D^\prime_\alpha$ derivative.

The next step is to put the terms involving $\eta^{\mu\nu}/a a'$ and
$\overline{\eta}^{\mu\nu}/a a'$ in (\ref{oldway:3}) into a covariant 
form using relations (\ref{dSeta}-\ref{dSetabar}). Upon inserting these 
relations into (\ref{oldway:3}), and then comparing with Eq.
(\ref{newway2}), we find,
\begin{eqnarray}
 f_1 &=& \frac{1}{2H^4}( \widehat{F} + \widehat{G} )
\;,\qquad f_2 = 0
\;,
\nonumber\\
 f_3 &=& \frac{1}{H^4}( \widehat{F} + \widehat{G} ) =  \tilde f_3
\;,
\nonumber\\
 f_4 &=& \frac{1}{H^4}(-y \widehat{F} +(2\!-\!y) \widehat{G})
\;,\qquad  f_5 = -\frac{1}{H^4}( \widehat{F} + \widehat{G})
\;.
\label{fi vs F and G}
\end{eqnarray}

The third step is to insert relations~(\ref{fi vs F and G}) into
the master structure functions Eqs.~(\ref{Phi1}--\ref{Psi5}) for
each of the five structure functions $f_i$, and then summing to
find,
\begin{eqnarray}
  \Phi & = & \sum_{i=1}^5 \Phi_i = -\frac{1}{2}\partial_y F
\;,
\label{master structure function Phi}
\\
  \Psi & = & \sum_{i=1}^5\Psi_i =
\frac{1}{2}[(D\!-\!1)+y\partial_y+2\partial_u]\partial_y F +\partial_y^2G
\nonumber\\
       &=& - [(D\!-\!1)-(2\!-\!y)\partial_y+2\partial_u]\Phi +\partial_y^2(F+G) 
\;.
\label{master structure function Psi}
\end{eqnarray}
This implies that we can express the vacuum polarization tensor in terms
of the two structure functions $f_1$ and $f_2$ as in Eq.~(\ref{newway}).
Indeed, with the help of Eqs.~(\ref{Phi1}--\ref{Psi2}) we get:
\begin{eqnarray}
 \Phi_1+\Phi_2&=&\frac{H^4}{4}\big(\partial_u^2-\partial_v^2\big)g_1
\;,\qquad g_1 = 2f_1 +(2\!-\!y)f_2
\nonumber\\
 \Psi_1+\Psi_2&=&\frac{H^4}{4}
              \big\{
                  [(D\!-\!1)-(2\!-\!y)\partial_y]g_1+4\partial_y f_2
              \big\}
\;.
\label{Phi and Psi 1+2}
\end{eqnarray} 
These equations can be solved for $g_1$ and $f_{2,1}$ in terms of 
$\Phi\equiv \Phi_1+\Phi_2$ and $\Psi\equiv \Psi_1+\Psi_2$.
Using the `light-cone' coordinates, $x_\pm=(u\pm v)/2$
($x_+=\ln(a), x_-=\ln(a^\prime)$), and requiring interchange
symmetry gives,
\begin{eqnarray}
 g_1(y,u,v)&=&\int_0^{x_+} \!\!\!\!\! d\bar x_+ \! \int_0^{x_-} 
\!\!\!\!\! d\bar x_-
                \Psi(y,\bar x_+ \!+\! \bar x_-,\bar x_+ \!-\! \bar x_-)
\nonumber\\
    && \hskip 2.2cm +\,\alpha(y,x_+) + \alpha(y,x_-)
\;,
\nonumber
\\
 f_2 &=& \frac{2\!-\!y}{4}g_1
     +\int \! dy^\prime\Big[\frac{1}{H^4}\Phi(y^\prime,u,v) 
                        - \frac{D\!-\!2}{4}g_1(y^\prime,u,v) \Big]
     +\beta(u,v)
\;,
\nonumber\\
 f_1 &=& \frac12[g_1-(2\!-\!y)f_2]
\;,
\label{g1 f1 and f2}
\end{eqnarray}
where $\alpha(y,x_\pm)$ and $\beta(u,v)$ are arbitrary functions,
and $\Psi$ and $\Phi$ are given in 
Eqs.~(\ref{master structure function Phi}--\ref{master structure function Psi}).
Together with Eqs.~(\ref{F G: rescaling}), these expressions
provide the desired procedure for transforming the vacuum polarization 
tensor from the form given in~(\ref{oldway}) 
into~(\ref{newway}--\ref{vacpol:dS}), and thus constitute
the main result of this paper.

 In the subsequent section we show that these transformations
indeed work for the vacuum polarization induced by the one loop 
vacuum fluctuations of a minimally coupled massless scalar 
on de Sitter background, providing a nontrivial check 
of Eqs.~(\ref{master structure function Phi}--\ref{g1 f1 and f2}). 


\section{Correspondence}
\label{Correspondence}

 Here we consider the vacuum polarization induced by 
one loop vacuum fluctuations of a massless minimally coupled scalar 
on de Sitter background. 

 If we call the scalar propagator $i\Delta(x;x')$ then
the one loop contribution to the vacuum polarization is of the form,
\begin{eqnarray}
\label{orig_vacpol} \lefteqn{i \Bigl[\mbox{}^\mu \Pi^\nu \Bigr](x;x')
= -2 i e^2 \sqrt{-g} \, g^{\mu\nu} i \Delta(x;x) \delta^D(x \!-\! x')
} \nonumber \\
& & \hspace{-.5cm} + 2 e^2 \sqrt{-g} \, g^{\mu\rho} \sqrt{-g'} \,
g'^{\nu\sigma} \Bigl[\partial_\rho i \Delta(x;x')
\partial'_\sigma i \Delta(x;x') \!-\! i \Delta(x;x')
\partial_\rho \partial'_\sigma i \Delta(x;x') \Bigr] \nonumber \\
& & \hspace{3.5cm} + i \delta Z \partial_{\rho} \Bigl[ \sqrt{-g}
\Bigl( g^{\mu\nu} g^{\rho\sigma} \!-\! g^{\mu\sigma} g^{\nu\rho}
\Bigr) \partial_{\sigma} \delta^D(x-x') \Bigr] \; , \qquad
\end{eqnarray}
were $\delta Z$ is a counterterm, and $e$ is the electromagnetic coupling
($e^2/(4\pi)\equiv \alpha_e \simeq 1/137$).
For the special case of the massless, minimally coupled scalar the
propagator obeys
\begin{equation}
\sqrt{-g} \, \square i \Delta_A(x;x') = i \delta^D(x-x')\; ,
\label{Apropeqn}
\end{equation}
where $\square \equiv (-g)^{-\frac12} \partial_{\mu} (\sqrt{-g} \,
g^{\mu\nu} \partial_{\nu})$ is the covariant d'Alembertian as it acts on 
a (bi-)scalar function. The unique solution 
that preserves the symmetries of the cosmological background 
(homogeneity, isotropy and spatial flatness)
can be written as a sum of a de Sitter invariant part
and a de Sitter breaking part ($\propto u=\ln(aa^\prime)$)  
as \cite{Onemli:2002hr,Onemli:2004mb}:
\begin{equation}
i \Delta_A(x;x') = A\bigl(y(x;x')\bigr) + k u 
\; ,
\end{equation}
where the constant $k$ is
\begin{equation}
k = \frac{H^{D-2}}{(4\pi)^{\frac{D}2}} \frac{\Gamma(D
\!-\!1)}{\Gamma(\frac{D}{2})} \; 
\label{k}
\end{equation}
and the de Sitter invariant part of the propagator $A(y)$ obeys -- 
as a consequence of the propagator equation (\ref{Apropeqn}) --
the equation
\begin{equation}
(4 y \!-\! y^2) A''(y) + D (2 \!-\! y) A'(y) = (D \!-\! 1) k \; .
\label{Aeqn}
\end{equation}
The solution of this equation
 can be written as a combination of two Gauss' hypergeometric
functions. Upon inserting these into~(\ref{orig_vacpol}),
one obtains for the vacuum polarization 
\begin{eqnarray}
\!i \bigl[\mbox{}^\mu \Pi^\nu \bigr](x;x') 
\!\!& = &\!\! 2 e^2 \sqrt{-g} \, g^{\mu\rho} \sqrt{-g'} \,
g'^{\nu\sigma} \Bigl\{ \partial_\rho y\partial'_\sigma y \!\times\!
\Bigl(A'^2 \!-\! A A'' \!-\! k u A'' \Bigr)
\qquad
\label{old_vacpol}
 \\
& & \hspace{-2.5cm} - \partial_\rho \partial'_\sigma y \!\times\!
\Bigl(A A' \!+\! k u A' \Bigr) \!+\! \partial_\rho u
\partial'_\sigma y \times k A' \!+\! \partial_\rho y
\partial'_\sigma u \!\times \! k A' \!+\! \partial_\rho u
\partial'_\sigma u \!\times\! k^2 \Bigr\} \nonumber 
\end{eqnarray}
plus local terms $\propto \delta^D(x-x^\prime)$, which are not 
relevant for this consideration (see {\it e.g.}
 Ref.~\cite{Leonard:2012si}).

 In order to establish correspondence, it suffices to
consider two independent components of 
$i \bigl[\mbox{}^\mu \Pi^\nu \bigr]$ in ~(\ref{old_vacpol}). 
It turns out that taking the $00$ and $ij$ ($i\neq j$) components
of~(\ref{old_vacpol}) leads to rather simple expressions,
\begin{eqnarray}
i \bigl[\mbox{}^0 \Pi^0 \bigr](x;x') 
\!\!& = &\!\! 2 e^2 \sqrt{-g} \, \sqrt{-g'} \,
 \frac{H^2}{aa^\prime}
\label{Pi 00}
\\
&&\times
  \Bigl\{ \big[4+(2\!-\!y)^2-4(2\!-\!y)\cosh(v)\big]
                 \bigl(A'^2 \!-\! A A'' \!-\! k u A'' \bigr)
\quad
\nonumber \\
& & \hspace{-3.2cm} +\,\big[(2\!-\!y)-4\cosh(v)\big]
                        \bigl(A A' \!+\! k u A' \bigr)
 \!+\!   \big[-2(2\!-\!y)+4\cosh(v)\big]\times k A' 
   \!+\!  k^2 \Bigr\} 
\nonumber \\
\!i \bigl[\mbox{}^i \Pi^j \bigr](x;x')|_{i\neq j} 
\!\!\!& = &\!\!\! 2 e^2 \sqrt{-g} \, \sqrt{-g'} \,
 \frac{H^2}{aa^\prime}
  \Bigl\{ \big[\!-\!4a\Delta x^ia^\prime\Delta x^j\big]
                 \bigl(A'^2 \!-\! A A'' \!-\! k u A'' \bigr)
\,.
\nonumber\\
\label{Pi ij}
\end{eqnarray}
 On the other hand, from Eq.~(\ref{oldway}) we read off, 
\begin{eqnarray}
i \bigl[\mbox{}^0 \Pi^0 \bigr]
\!\!& = &\!\! -\nabla^\prime\cdot\nabla F 
\nonumber
\\
    \!\!& = &\!\! 2\sqrt{-g}\sqrt{-g^\prime}\frac{H^2}{aa^\prime }
   \Big[(D\!-\!1)-2(2\!-\!y)\partial_y 
              +4\cosh(v)\partial_y\Big]\partial_y \widehat{F}
\label{Pi 00:FH} \\
\!i \bigl[\mbox{}^i \Pi^j \bigr]_{i\neq j} 
\!\!& = &\!\! -\partial^\prime_i\partial_j (F + G)
\nonumber
\\
    \!\!&=&\!\!  \sqrt{-g} \, \sqrt{-g'} \,
 \frac{H^2}{aa^\prime}
  \Bigl\{ \big[4a\Delta x^ia^\prime\Delta x^j\big]\partial_y^2
( \widehat{F} + \widehat{G})\Bigl\}
\,.
\label{Pi ij:FG}
\end{eqnarray}
We see from Eq.~(\ref{Pi 00}) that $i \bigl[\mbox{}^0 \Pi^0 \bigr]$
contains dependencies in the form: (a) a function of $y$, 
(b) a function of $y$ times $u$ and (c) a function of $y$ times $\cosh(v)$.
On the other hand, inspecting~(\ref{Pi 00:FH})
and assuming that $\widehat{F}$ does not depend on $v$ 
(this in fact follows from the spatial diagonal components 
$i \bigl[\mbox{}^i \Pi^i \bigr]$)
tells us that the following equations must be separately satisfied,
\begin{eqnarray}
 e^2\!\!\!\!&&\!\!\!\!\Big\{
     \!-\!2(2\!-\!y)\bigl(A^{\prime 2} \!-\! A A^{\prime\prime} 
     \!-\! k u A^{\prime\prime} \bigr)
     -2\bigl(A \!+\! k u \bigr)A^\prime + 2 kA^\prime 
    \Big\}
   = 2 \partial_y^2 \widehat{F} 
\nonumber\\
 e^2\!\!\!\!&&\!\!\!\!\Big\{
     [4+(2\!-\!y)^2]\bigl(A^{\prime 2} \!-\! A A^{\prime\prime} 
     \!-\! k u A^{\prime\prime} \bigr) 
\nonumber\\
   &&\!\!\!
    +\,2(2\!-\!y)\bigl(A \!+\! k u \bigr)A^\prime 
     - 2(2\!-\!y) kA^\prime +k^2 
    \Big\}
     = \big[(D\!-\!1)-2(2\!-\!y)\partial_y\big] \partial_y \widehat{F} 
\,.\quad
\nonumber
\end{eqnarray}
 Multiplying the first equation by $(2\!-\!y)$ and adding the second results 
in,
\begin{eqnarray}
 e^2\Big\{
     (4y\!-\!y^2)A^{\prime 2} \!+\!(D\!-\!1)(2\!-\!y)(A+ku)A^\prime 
     \!-\! k(D\!-\!1)(A+ku)+k^2 \bigr)\Big\}&& 
\nonumber\\
     &&\hskip -3cm
   =\, (D\!-\!1) \partial_y \widehat{F}
\,, \quad
\nonumber
\end{eqnarray}
where we also made use of Eq.~(\ref{Aeqn}).
Now, multiplying this by $-1/[2(D\!-\!1)]$
and separating the $u$-dependent terms yields
\begin{eqnarray}
 -\frac12  \partial_y \widehat{F} 
\!&=&\! 
 -\frac{e^2}{2(D\!-\!1)}\Big\{(D\!-\!1)\big[(2\!-\!y)A^\prime-k\big]A 
         \!+\!(4y\!-\!y^2)A^{\prime 2}  \!+\! k^2 \Big\}
\nonumber\\
    && -\,\frac{e^2k}{2}\big[(2\!-\!y)A^\prime\!-\!k\big] u
   =  \Phi
\,,
\label{Phi for MMCS}
\end{eqnarray}
where the last equality follows from ~(\ref{master structure function Phi}).
This equation agrees with Eq.~(44) in Ref.~\cite{Leonard:2012si},
establishing the first part of the correspondence for the minimally
coupled massless scalar field.

 Next we consider Eqs.~(\ref{Pi ij}) and~(\ref{Pi ij:FG}), which give us 
information about the second structure function $G$,
\begin{eqnarray}
-2e^2  \bigl(A'^2 \!-\! A A'' \!-\! k u A'' \bigr)  = \partial_y^2
(\widehat{F} + \widehat{G})
\,.
\label{Pi ij:2}
\end{eqnarray}
In order to get the second master structure 
function~(\ref{master structure function Phi}), we 
add 
\begin{eqnarray}
-[(D\!-\!1)-(2\!-\!y)\partial_y+2\partial_u]\Phi
   \!\!&=&\!\! 2e^2  \bigl(A'^2 \!-\! A A'' \!-\! k u A'' \bigr)
\nonumber\\
   \!\!&&\!\!  +\,\frac{e^2k}{2}\big[(2\!-\!y)A^\prime\!-\!k\big]
\,
\end{eqnarray}
to both sides of Eq.~(\ref{Pi ij:2}). This procedure results in 
\begin{equation}
 \Psi = \frac{e^2k}{2}\big[(2\!-\!y)A^\prime\!-\!k\big]
\,,
\label{Psi: MCMS}
\end{equation}
and agrees with Eq.~(45) of Ref.~\cite{Leonard:2012si}.
This establishes the second part of the correspondence for 
the minimally coupled massless scalar field,
and completes our (nontrivial) check of 
the main results~(\ref{master structure function Phi}--\ref{g1 f1 and f2}).


\section{Discussion}
\label{Discussion}

We have established how to transform from the non-covariant
form (\ref{oldway}) of the vacuum polarization tensor used in 
Refs.~\cite{Prokopec:2002jn,Prokopec:2002uw,Prokopec:2003iu,Prokopec:2003tm}
to the covariant form (\ref{newway}) introduced in our recent 
work~\cite{Leonard:2012si}. Our main results are 
the transformation 
formulae~(\ref{master structure function Phi}--\ref{g1 f1 and f2}),
based on which one can effortlessly move from one representation
of the vacuum polarization to another. Our results can be useful for studies 
of the one loop vacuum polarization induced by gravitons, scalars and 
fermions on a locally de Sitter background~\cite{Leonard:2012}. 
In particular, it provides a way of characterizing the irreducible
effect on photons of de Sitter breaking in the gravitational sector. 
An important technical result of some importance is the
improved procedure (\ref{Pi 00:FH}-\ref{Pi ij:FG}) for
extracting the noncovariant structure functions $F(x;x)$ and
$G(x;x')$.

Our results might be useful for understanding the structure of the 
photon vacuum polarization tensor on arbitrary background geometries. 
Namely, on de Sitter background the vacuum polarization is given 
in terms of two bi-scalar structure functions which generally violate
de Sitter symmetry. In the special case, when the de Sitter symmetry 
is unbroken, one (de Sitter invariant) structure function suffices 
to fully specify the vacuum polarization tensor.
It is tempting to try generalizing our results~(\ref{newway})
to generic backgrounds by replacing the de Sitter invariant length
$y(x;x^\prime)$ by the appropriate geodesic distance $\ell(x;x^\prime)$,
in which case~(\ref{newway}) would become
\begin{equation}
\bigl[\mbox{}_{\mu\rho} T_{\nu\sigma} \bigr](x;x^\prime)
= \partial_\mu \partial'_{[\nu} \ell^2 \, \partial'_{\sigma]} \partial_\rho \ell^2
\!\times\! f_1 + \partial_{[\mu} \ell^2 \, \partial_{\rho]} \partial'_{[\nu} \ell^2 \,
\partial'_{\sigma]} \ell^2 \!\times\! f_2
\,.
\label{TensorT:ell}
\end{equation}
One would think there must be at least two structure functions
in a general metric because photons have two polarizations and
birefringence is known to occur in some backgrounds \cite{Hu:1997iu}.
On the other hand, there cannot be more than six structure functions 
because that is the number of independent components of a transverse
bi-vector density. It would be interesting to see if tighter bounds 
can be derived.

\vskip 1cm

\centerline{\bf Acknowledgements}

This work was partially supported by NSF grant PHY-1205591, and by 
the Institute for Fundamental Theory at the University of Florida.

\end{document}